\documentclass[prl,showkeys,preprintnumbers,showpacs,superscriptaddress,twocolumn]{revtex4}
\pdfoutput=1
\usepackage{graphicx}% Include figure files
\usepackage{dcolumn}% Align table columns on decimal point
\usepackage{pslatex}% use standard PS fonts & reduce PS file
\usepackage{amsmath}% for advanced Eq. numering
\usepackage{amssymb}% for advanced Eq. numering
\usepackage{bm}% bold math
\usepackage{color} % textcolor
\usepackage{nicefrac}
\usepackage{ulem}

\newcommand{\gapprox}{{\scriptscriptstyle\stackrel{>}{\sim}}}
\newcommand{\lapprox}{{\scriptscriptstyle\stackrel{<}{\sim}}}

\begin{document}

%\preprint{G\"{u}rlich {\em et al.} -- LTSEM on zigzag JJs -- corrected proofs version, August 2009}

\title{Imaging of order parameter induced $\pi$ phase shifts in
cuprate superconductors by low-temperature scanning electron
microscopy}

\author{Christian G\"{u}rlich}
\author{Edward Goldobin}
\author{Rainer Straub}
\author{Dietmar Doenitz}
\affiliation{Physikalisches Institut - Experimentalphysik II and
Center for Collective Quantum Phenomena, Universit\"{a}t T\"{u}bingen, Auf
der Morgenstelle 14, D-72076 T\"{u}bingen, Germany}
\author{Ariando}
\affiliation{Faculty of Science and Technology and MESA$^+$ Institute
for Nanotechnology, University of Twente, P.O. Box 217, 7500 AE
Enschede, The Netherlands}
\affiliation{Nanocore and Department of Physics, Faculty of Science,
National University of Singapore, Singapore 117542, Singapore}
\author{Henk-Jan H. Smilde}
\author{Hans Hilgenkamp}
\affiliation{Faculty of Science and Technology and MESA$^+$ Institute
for Nanotechnology, University of Twente, P.O. Box 217, 7500 AE
Enschede, The Netherlands}
\author{Reinhold Kleiner}
\author{Dieter Koelle}
\email{koelle@uni-tuebingen.de}
\affiliation{Physikalisches Institut - Experimentalphysik II and
Center for Collective Quantum Phenomena, Universit\"{a}t T\"{u}bingen, Auf
der Morgenstelle 14, D-72076 T\"{u}bingen, Germany}



\date{\today}

\begin{abstract}

Low-temperature scanning electron microscopy (LTSEM) has been used to
image the supercurrent distribution in ramp-type Josephson junctions
between Nb and either the electron-doped cuprate
Nd$_{2-x}$Ce$_x$CuO$_{4-y}$ or the hole-doped cuprate
YBa$_2$Cu$_3$O$_7$.
For zigzag-shaped devices in the short junction limit the critical current is
strongly suppressed at zero applied magnetic field.
The LTSEM images show, that this is due to the Josephson current
counterflow in neighboring 0 and $\pi$ facets, which is induced by
the $d_{x^2-y^2}$ order parameter in the cuprates.
Thus, LTSEM provides imaging of the sign change of the superconducting order
parameter, which can also be applied to other types of Josephson junctions.

\end{abstract}

\maketitle

One of the most controversial topics on high-$T_c$ cuprate superconductors has
been the determination of their order parameter symmetry (OPS).
A myriad of experiments have been performed, indicating a predominant
$d_{x^2-y^2}$ OPS, which implies important consequences for the microscopic
mechanism of Cooper pairing in these materials.
Obviously, it was quite difficult to identify an unambiguous experiment for the
determination of the cuprate OPS.
Among the most convincing experiments is the observation of
half-integer magnetic flux quanta in tricrystal grain boundary
Josephson junctions (JJs) by scanning SQUID microscopy
\cite{Tsuei00}.
These experiments, and related integral measurements of critical
current $I_c$ vs applied magnetic field $B$, rely on the difference
$\pi$ of the phase of the order parameter between orthogonal
directions in $(k_x,k_y)$-space, which can be detected by
interferometer-type configurations, such as corner junctions
\cite{VanHarlingen95}, tricrystal rings and long JJs
\cite{Tsuei94,Kirtley95c,Kirtley99a,Kirtley96,Tsuei00a,Sugimoto02a},
and dc $\pi$ SQUIDs \cite{Wollman93,Mathai95,Schulz00,Chesca03}, or
by the angular dependence of $I_c$ in biepitaxial JJs
\cite{Lombardi02a}.
High-quality hybrid ramp-type JJs, combining an $s$-wave superconductor (Nb)
with either the hole-doped cuprate YBa$_2$Cu$_3$O$_{7-\delta}$ (YBCO)
\cite{Smilde02a,Smilde02,Hilgenkamp03} or the electron-doped cuprate
Nd$_{2-x}$Ce$_x$CuO$_{4-y}$ (NCCO) \cite{Ariando05} have also been realized.
Arranging such JJs in a zigzag geometry with the facets oriented along the $a$-
and $b$-axis of the cuprate, one obtains alternating facets of 0 and $\pi$ JJs
\cite{Smilde02,Ariando05}.
$\pi$ JJs \cite{Bulaevskii77} have negative $I_c$, i.~e., $j_s=-j_c
\sin{\phi}=j_c \sin(\phi+\pi)$, instead of $j_s=j_c \sin{\phi}$, where $j_s$ is
the supercurrent density; $j_c>0$ is the maximum supercurrent density, and
$\phi$ is the Josephson phase.
Realizations include JJs with magnetic barriers
\cite{Ryazanov01,Kontos02,Blum02,Weides06,Vavra06}, geometric constrictions in
$d$-wave superconductors \cite{Gumann07}, Nb JJs with a mesoscopic Au control
channel \cite{Baselmans99}, Al JJs with a controllable quantum dot in a InAs
nanowire \cite{VanDam06}, and gate-controlled carbon nanotube JJs
\cite{Cleuziou06}.
JJs containing both, 0- and $\pi$-parts have also been realized using
ferromagnetic barriers \cite{DellaRocca05,Frolov06,Weides06a} or current
injectors \cite{Goldobin04}.

A striking property of $s$-$d$-wave zigzag JJs in the {\it long JJ limit}
(facet length $a\gapprox 4\lambda_J$) is the spontaneous generation of magnetic
flux $\pm\Phi_0/2$, i.e.~a semifluxon at each corner of the zigzag
($\Phi_0=h/2e$ is the magnetic flux quantum and $\lambda_J\propto j_c^{-1/2}$
the Josephson penetration depth).
The presence of semifluxons in such devices was demonstrated
\cite{Hilgenkamp03} by scanning SQUID microscopy.
In the {\it short JJ limit} (neglecting self-field effects), for a JJ with $N$
facets, the supercurrent density in the $n^{th}$ facet can be described as
\cite{Smilde02}
\begin{equation}
j_s(\tilde{x})=(-1)^nj_c(\tilde{x})\sin\{\phi_0+(2\pi\Phi_f/\Phi_0Na)\cdot \tilde{x}\}\quad.
\label{eq:js}
\end{equation}
Here, $\tilde{x}$ is the coordinate along the zigzag (with
$\tilde{x}=0$ at the JJ edge), and $\Phi_f$ is the magnetic flux per
facet.
As the prefactor $(-1)^n$ changes sign at every corner of the zigzag, as a
direct consequence of the $d$-wave OPS, $I_c(B)$ is not Fraunhofer-like;
instead, it has main maxima ($I_c^{max}=(\nicefrac{2}{\pi})Nj_c ha$ for
$j_c(\tilde{x})=\rm const.$) at finite field, corresponding to
$\Phi_f=\pm\Phi_0/2$ for even $N$, with junction area $h\cdot a$ per facet.
According to Eq.~(\ref{eq:js}), at such $\Phi_f$,
$j_s(\tilde{x})=j_c|\sin\nicefrac{\pi\tilde{x}}{a}|$ in each facet.
$I_c(B)$ at $B=0$ has a minimum (for even $N$) or a small local maximum (for
odd $N$).
In the case of homogeneous $j_c(\tilde{x})$, absence of self-field effects and
even $N$ one expects $I_c(0)=0$, due to a current distribution $j_s=(-1)^n
j_c$, and current reversal at each corner of the zigzag results in a quite
unusual $I_c(B)$ dependence \cite{Smilde02,Ariando05}, which provides strong
(indirect) evidence of the Josephson current counterflow as a direct
consequence of the sign change in the $d$-wave order parameter.

In this Letter we show that low-temperature scanning electron microscopy
(LTSEM) allows imaging of the supercurrent distribution in YBCO-Nb and NCCO-Nb
JJs, and we demonstrate Josephson current counterflow in 0- and $\pi$-facets in
zigzag shaped cuprate/Nb JJs at $B=0$.

We investigated hybrid ramp-type JJs with 150\,nm thick [001] YBCO or optimally
doped ($x$=0.15) NCCO bottom electrodes, grown epitaxially on [001] SrTiO$_3$
(STO) single-crystal substrates and covered by an STO film with thickness
100\,nm and 35\,nm, respectively.
After milling a shallow ramp (15$^\circ$--20$^\circ$) into the bilayers, an
epitaxial YBCO (6\,nm) or NCCO (12\,nm) interlayer was grown, followed by {\it
in-situ} deposition of a Au barrier layer of thickness $d_{\rm Au}$, and a Nb
layer (140-160\,nm) as a counter electrode \cite{Smilde02a,Smilde02,Ariando05}.
In total, we investigated four chips with identical layout.
Three chips contained YBCO-Nb JJs with $d_{\rm Au}$=14\,nm (chip Y1)
and 12\,nm (Y2 and Y3) in order to investigate samples with different
$j_c$, i.e.~different $\lambda_J$.
The chip N with the NCCO-Nb JJs had $d_{\rm Au}$=12\,nm.
Below, we show data from chips N, Y1 and Y2 for zigzag JJs with $N=8$ and
$a$=25\,$\mu$m (chip Y1) or $N$=10 and $a$=40\,$\mu$m (chip N) and for
reference single facet JJs ($a$=50\,$\mu$m), oriented along the $a,b$ axis of
the cuprate film (chips N and Y2).
The conversion from $B$ (normal to the substrate plane) to magnetic flux $\Phi$
in the JJ was done by comparing the measured $I_c(B)$ with $I_c(\Phi)$
calculated from Eq.~(3) in Ref.~[\onlinecite{Smilde02}].
Considering the idle region (overlap of the Nb electrode on top of the
cuprate), we can only give a rough estimate on an upper limit for the
normalized JJ length $Na/\lambda_J\lapprox 2$, i.e.~all devices are expected to
be in the short JJ limit.
Regarding further electric transport properties of our samples, see
Refs.~[\onlinecite{Smilde02a,Smilde02,Ariando05,Chesca06}].

For imaging by LTSEM, the sample was mounted on a He cryostage and operated at
a temperature $T\approx$5-6\,K.
The local perturbation by the focused electron beam ($e$-beam) centered at the
position ($x_0,y_0$) on the sample surface in the $(x,y)$ plane induces an
increase in temperature $\delta T(x-x_0,y-y_0)$ on a lateral length scale of
$\approx$1-3\,$\mu$m, which determines the spatial resolution of this imaging
technique.
The maximum local increase in temperature $\Delta T$ is typically $<$1\,K, and
can be adjusted by the $e$-beam voltage $V_b$ and beam current $I_b$
\cite{Clem80,Gross94}.
For the LTSEM images shown below $V_b$=10\,kV and $I_b$=50\,pA-1\,nA.
$\delta T$ results in a local reduction of $j_c(T)$ and a concomitant change of
the overall $I_c$ of the JJ.
It has been shown theoretically \cite{Chang84,Chang85} and experimentally
\cite{Bosch85,Bosch87} that this effect can be used to image the spatial
distribution of the supercurrent density $j_s(\tilde{x})$ (at $I=I_c$,
convoluted with the $\delta T$ profile) along a short JJ by recording the
beam-induced change $\delta I_c(\tilde{x})\propto j_s(\tilde{x})$ of the
overall critical current as a function of the beam coordinate $\tilde{x}$,
during scanning along the JJ.
For simplicity, rather than detecting $\delta I_c$, we current bias the JJ
slightly above $I_c$ (typically at a voltage $V$ of a few $\mu$V) and detect
the beam-induced voltage change $\delta V$ \cite{Gross94}.
Assuming a constant differential resistance $R_d$ yields $\delta
V(\tilde{x})= -R_d\delta I_c(\tilde{x})\propto j_s(\tilde{x})$.
To improve the signal-to-noise ratio, we modulate the $e$-beam at 5\,kHz
(6.6\,kHz) and lock-in detect the voltage response from the YBCO(NCCO)-Nb JJs.

\begin{figure}[b]
\center{\includegraphics[width=8.5cm]{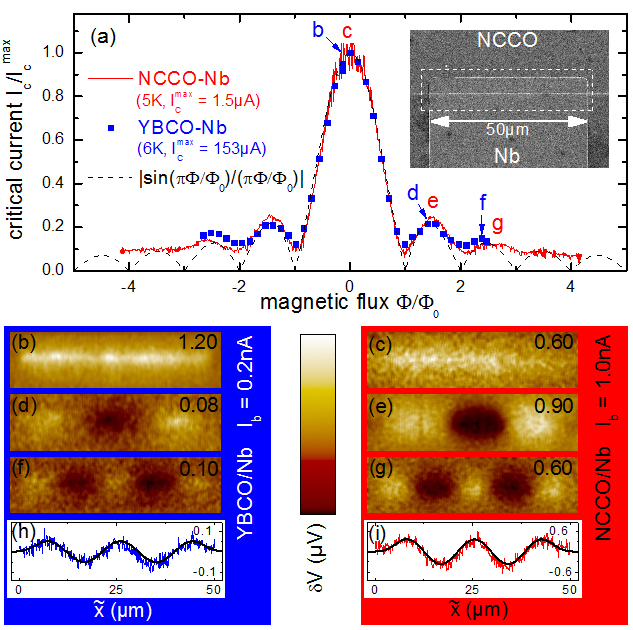}}
\caption{
(color online).
Single facet NCCO- and YBCO-Nb JJs:
(a)
Normalized critical current $I_c/I_c^{max}$ vs magnetic flux $\Phi/\Phi_0$.
Labels (b)--(g) indicate working points for LTSEM images below.
Inset: image of NCCO-Nb JJ;
dashed frame indicates size and position of LTSEM
images (b) -- (g).
Numbers indicate full range $|\delta V_{max}|$ (in $\mu$V) of the
scale bar (symmetric about $\delta V=0$).
(h),(i): linescans $\delta V(\tilde{x})$ at $\Phi/\Phi_0=5/2$
along the JJs, respectively, from images (f) and (g),
and calculated current density $j_s(\tilde{x})/j_c$ (solid black lines).}
\label{RefJJs}
\end{figure}

In order to characterize the quality of our devices and to
demonstrate imaging of the current distribution by LTSEM, we first
present results from the YBCO-Nb and NCCO-Nb single facet ($N$=1,
$a$=50\,$\mu$m) reference JJs.
The inset in Fig.~\ref{RefJJs}(a) shows an SEM image of the NCCO-Nb
JJ.
Figure \ref{RefJJs}(a) shows normalized critical current
$I_c/I_c^{\rm max}$ vs applied magnetic flux $\Phi=N\Phi_f$.
Fraunhofer-like $I_c$ oscillations are clearly visible, although
deviations from the ideal characteristic (dashed line) are obviously
present.
Those deviations are probably mainly due to the finite voltage
criterion for the detection of $I_c$, however also indicate
inhomogeneities in $j_c(\tilde{x})$.

%%%%%%%%%%%%% Fig. 1 %%%%%%%%%%%%%%%%%%%%%%%%%%%%%%%%%%%%%%%%%%%%%%%%%%
%%%%%%%%%%%%% Fig. 1 %%%%%%%%%%%%%%%%%%%%%%%%%%%%%%%%%%%%%%%%%%%%%%%%%%

Figures \ref{RefJJs}(b)-\ref{RefJJs}(g) show LTSEM images $\delta V(x_0,y_0)$
for both reference JJs (left: YBCO-Nb; right: NCCO-Nb) taken at different
values of $\Phi$ as indicated in Fig.~\ref{RefJJs}(a).
At $\Phi=0$ [main $I_c$ maximum; graphs (b) and (c)] the voltage
signals at $y_0=0$ ($\tilde{x}$-axis) are positive along the entire
length of both JJs.
At $\Phi=0$, for $N=1$ one finds from Eq.~(\ref{eq:js}) that the supercurrent
density at $I_c$ is $j_s(\tilde{x})=j_c(\tilde{x})$, and hence $\delta
V(\tilde{x})\propto j_c(\tilde{x})$, i.~e., the variation in $\delta
V(\tilde{x})$ along the JJ directly yields the variation of $j_c(\tilde{x})$.
The observed $\delta V(\tilde{x})$ clearly indicates $j_c$ inhomogeneities
along the JJs, which are most likely due to variations in the quality of the
interface and in the thickness of the Au barrier layer.
For the YBCO-Nb JJ, we find a maximum variation in $j_c(\tilde{x})$ of $\pm
15\,\%$.
For the NCCO-Nb JJ we observe a steplike decrease of $j_c(\tilde{x})$ at
$\tilde{x}\approx 35\mu$m by $\sim$30\,\%.

The second row of LTSEM images [graphs (d) and (e)] are taken at the first side
maximum in $I_c(\Phi)$, i.~e.~at $\Phi=\frac{3}{2}\Phi_0$ for which one expects
a sinusoidal variation of the supercurrent density
$j_s(\tilde{x})=j_c(\tilde{x})\sin{(3\pi \tilde{x}/a)}$ with 3/2 wavelengths.
This behavior is well confirmed by the LTSEM images.
The lowest row of LTSEM images [graphs (f) and (g)] for
$\Phi=\frac{5}{2}\Phi_0$, i.~e.~taken at the second side maximum in $I_c(\Phi)$
again clearly shows the expected oscillation with 5/2 wavelengths.
The graphs (h) and (i) in Fig.~\ref{RefJJs} show linescans taken from
the corresponding LTSEM images (f) and (g), together with the
calculated normalized current density distribution
$j_s(\tilde{x})/j_c$, which was convoluted with a Gaussian
beam-induced temperature profile $e^{-(x-\tilde{x})^2/2\sigma^2}$
with $\sigma=2.5\,\mu$m.
The excellent agreement between the measured voltage signals and
calculated current distribution clearly demonstrates that we indeed
image the supercurrent density distribution along the JJs.

%%%%%%%%%%%%% Fig. 2 %%%%%%%%%%%%%%%%%%%%%%%%%%%%%%%%%%%%%%%%%%%%%%%%%%
\begin{figure}[b]
\center{\includegraphics[width=8.5cm,clip]{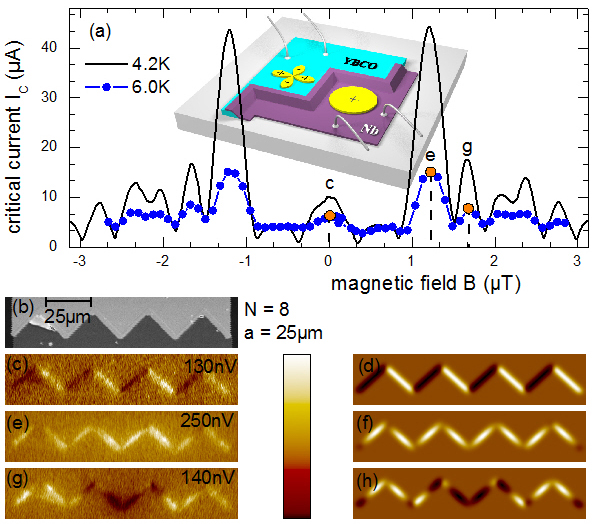}}
\caption{
(color online).
YBCO-Nb zigzag JJ:
(a) $I_c(B)$ patterns;
inset: sketch of zigzag-shaped ramp JJ.
(b) Surface image and (c), (e), (g) corresponding LTSEM images ($I_b=50\,$pA)
taken at different values for $B$ as indicated in (a);
(d), (f), (h) show corresponding calculated images of current distribution
along the zigzag.
\label{Yz}}
\end{figure}
%%%%%%%%%%%%% Fig. 2 %%%%%%%%%%%%%%%%%%%%%%%%%%%%%%%%%%%%%%%%%%%%%%%%%%

In the following, we present results on the zigzag JJs [c.f.~a schematic view
in the inset of Fig.~\ref{Yz}(a)], starting with the YBCO-Nb JJ ($N$=8,
$a$=25\,$\mu$m); Fig.~\ref{Yz}(b) shows an SEM image of this device.
Figure \ref{Yz}(a) shows $I_c(B)$ measured on the LTSEM cryostage at
$T$$\approx$6\,K (dots) and in a liquid He cryostat at $T$=4.2\,K (solid line).
As expected for an array of $0$-$\pi$ facets, $I_c(B)$ shows main maxima at
finite field ($B_{max}$=1.1\,$\mu$T) and only a small central maximum at $B$=0.
Because of the higher temperature of the LTSEM cryostage, the $I_c$ values are
reduced, as compared to the 4.2\,K data and the $I_c$ oscillations are washed
out.
Nevertheless, almost all maxima and minima in $I_c(B)$ still show up
at $T\approx 6\,$K.

For each point of the $I_c(B)$ dependence at 6\,K in Fig.~\ref{Yz}(a) LTSEM
images were recorded.
Figures \ref{Yz}(c), \ref{Yz}(e) and \ref{Yz}(g) [left row] show images taken
at three values of $B$ [as labeled in graph (a)], namely at the small maximum
in $I_c(B)$ at $B$=0 (c), at the main maximum in $I_c(B)$ (e), and at the next
side maximum in $I_c(B)$ (g).
To the right of each LTSEM image, we show the corresponding image
$j_s(x_0,y_0)$ of the supercurrent density distribution (normalized to a
spatially homogeneous $j_c$) which was calculated as follows:
The 1D distribution $j_s(\tilde{x})$ along a zigzag line in the
$x$-$y$ plane  was calculated numerically from Eq.~(\ref{eq:js}), and
all the points $(x,y)$ outside the zigzag line were set to $j_s=0$.
The resulting 2D $j_s(x,y)$ distribution was then convoluted with a Gaussian
profile, i.~e.,
\newcommand{\xmin}{\ensuremath{x_\mathrm{min}}}
\newcommand{\xmax}{\ensuremath{x_\mathrm{max}}}
\newcommand{\ymin}{\ensuremath{y_\mathrm{min}}}
\newcommand{\ymax}{\ensuremath{y_\mathrm{max}}}
$j_s(x_0,y_0) = \int_{\xmin}^{\xmax}
\int_{\ymin}^{\ymax}j_s(x,y)\exp\{-r^2/2\sigma^2\}\,dx\,dy$,
with $r^2=(x-x_0)^2+(y-y_0)^2$ and $\sigma = 2.5\,\mathrm{\mu m}$, and plotted
in Figs.~\ref{Yz}(d), \ref{Yz}(f), \ref{Yz}(h).
The calculated images are in good qualitative agreement with the
LTSEM images.
As the main result, Fig.~\ref{Yz}(c) clearly shows the alternating
sign of supercurrent flow across neighboring facets at $B$=0.
Thus, the LTSEM image provides a direct proof of the existence of $0$
and $\pi$ facets in the zigzag JJ, due to the sign change of the
order parameter in the $d$-wave cuprate superconductor YBCO.
In contrast, Fig.~\ref{Yz}(e) taken at the main maximum in $I_c(B)$,
shows only positive voltage signals which are largest inside the
facets and which tend to zero at the corners.
This is in qualitative agreement with $j_s(\tilde{x})\propto|\sin{\pi
\tilde{x}/a}|$ as expected for a homogeneous zigzag JJ with $j_c$=const.
Quantitative differences as observed by the LTSEM voltage signals can
most likely be attributed to $j_c$ inhomogeneities along the zigzag
JJ, as such inhomogeneities have already been observed for the
YBCO-Nb reference JJ [c.~f.~Fig.~\ref{RefJJs}(b)].
The LTSEM image recorded at the next side maximum in $I_c(B)$
[Fig.~\ref{Yz}(g)] shows a polarity of the voltage signals (positive outside
and negative in the center) which is reminiscent of the behavior of the
reference JJs also biased at the first side maximum in $I_c(B)$
[c.~f.~Figs.~\ref{RefJJs}(d) and \ref{RefJJs}(e)].
Again, this is in qualitative agreement with the calculated $j_s(\tilde{x})$
for the zigzag JJ with homogeneous $j_c$ distribution [c.~f.~Fig.~\ref{Yz}(h)].

%%%%%%%%%%%%% Fig. 3 %%%%%%%%%%%%%%%%%%%%%%%%%%%%%%%%%%%%%%%%%%%%%%%%%%
\begin{figure}[b]
\center{\includegraphics[width=8.5cm,clip]{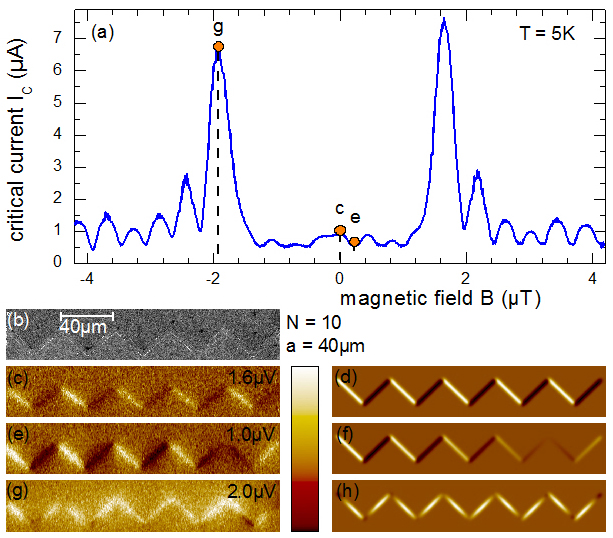}}
\caption{
(color online).
NCCO-Nb zigzag JJ:
(a) $I_c(B)$ pattern;
(b) surface image; (c), (e), (g) LTSEM voltage images ($I_b=1\,$nA)
taken at different values for $B$ as indicated in (a).
(d), (f), (h): corresponding calculated images of current distribution
along the zigzag.
\label{Nz}}
\end{figure}
%%%%%%%%%%%%% Fig. 3 %%%%%%%%%%%%%%%%%%%%%%%%%%%%%%%%%%%%%%%%%%%%%%%%%%

Finally, we demonstrate that similar results were obtained by imaging the
current distribution in the NCCO-Nb zigzag JJ ($N$=10, $a$=40\,$\mu$m);
c.~f.~the SEM image in Fig.~\ref{Nz}(b).
Figure \ref{Nz}(a) shows $I_c(B)$ measured on the LTSEM cryostage at
$T$$\approx$5\,K, which was almost identical to $I_c(B)$ measured in liquid He
at 4.2\,K.
As for the YBCO-Nb zigzag JJ, $I_c(B)$ has a small central $I_c$ maximum and
main $I_c$ maxima at finite field.
Figures \ref{Nz}(c), \ref{Nz}(e) and \ref{Nz}(g) show LTSEM images taken at
three values of $B$ [as labeled in graph (a)], namely at the small central
maximum in $I_c(B)$ at $B=0$ (c), at the ''dip`` in $I_c(B)$ close to $B$=0 (e)
and at the main maximum in $I_c(B)$ (g).
As in Fig.~\ref{Yz}, the corresponding calculated images (d), (f), (h) of
$j_s(x_0,y_0$) (with $\sigma$=2.5\,$\mu$m) are in qualitative agreement with
the LTSEM images.
Again, Fig.~\ref{Nz}(c) clearly shows the alternating sign of
supercurrent flow across neighboring facets at $B$=0.
This pattern remains almost unchanged in a small applied field [bias
point ''e`` in (a)] as shown in Fig.~\ref{Nz}(e).
Here the polarity of the LTSEM voltage signals for the two facets at
the right edge of the JJ changed.
Probably due to the $j_c$ inhomogeneity along the entire JJ this
state results in an even lower value of $I_c$ as compared to the
$I_c$ value at $B$=0.
At the main $I_c$ maximum, the LTSEM image in Fig.~\ref{Nz}(g) again
shows only positive voltage signals, as expected, and as discussed
above.

In conclusion, we have shown that low-temperature scanning electron
microscopy allows imaging of the supercurrent distribution in
cuprate-Nb hybrid ramp-type Josephson junctions.
LTSEM images recorded at $B=0$ show Josephson current counterflow.
This gives direct evidence of the presence of alternating $0$ and
$\pi$ facets in YBCO-Nb and NCCO-Nb zigzag junctions, which is due to
the sign change of the $d$-wave order parameter in the cuprate
superconductors involved in this study.
We note, that the same technique can also be applied to other systems which
produce $0$-$\pi$ Josephson junctions, e.g.~JJs with a ferromagnetic barrier.
Furthermore, this technique may also be applied to investigate the order
parameter symmetry in less studied superconducting materials, if they can be
combined with an $s$-wave superconductor to form hybrid Josephson junctions.

This work was supported by the DFG (Kl930/11), the FOM, the NWO and by the ESF
programme NES.

%\bibliography{guerlich09zigzag}
% Produces the bibliography via BibTeX
\bibliography{guerlich09zigzag-etal}
% -- with et al. for four and more co-authors

\end{document}